# Two-Dimensional Van der Waals Epitaxy Kinetics in a Three-Dimensional Perovskite Halide

Yiping Wang[1], Yunfeng Shi[1], Guoqing Xin[2], Jie Lian[2], Jian Shi[1*]

[1] Department of Materials Science and Engineering, Rensselaer Polytechnic Institute, Troy, NY, 12180

[2] Department of Mechanical, Aerospace, and Nuclear Engineering, Rensselaer Polytechnic Institute, Troy, NY, 12180

[*]Corresponding author: shij4@rpi.edu

**Abstract**

The exploration of emerging materials physics and prospective applications of two-dimensional materials greatly relies on the growth control of their thickness, phases, morphologies and film-substrate interactions. Though substantial progresses have been made for the development of two-dimensional films from conventional layered bulky materials, particular challenges remain on obtaining ultrathin, single crystalline, dislocation-free films from intrinsically non-Van der Waals-type three-dimensional materials. In this report, with the successful demonstration of single crystalline ultrathin large scale perovskite halide material, we reveal and identify the favorable role of weak Van der Waals film-substrate interaction on the nucleation and growth of the two-dimensional morphology out of non-layered materials compared to conventional epitaxy. We also show how the bonding nature of the three-dimensional material itself affects the kinetic energy landscape of ultrathin films growth. By studying the formation of fractal perovskites assisted with Monte Carlo simulations, we demonstrate that the competition between the Van der Waals diffusion and surface free energy of the perovskite leads to film thickening, suggesting extra strategies such as surface passivation may be needed for the growth of monolayer and a few layers films.

## 1. Introduction

Featured by the rich emerging physics[1] and technological promises[2], two-dimensional (2D) materials have attracted much attention from both academic and industrial communities. Recently, endeavors have been extended from traditional graphene research to other van der Waals (VDW) materials exemplified by layered transition metal dichalcogenides[3-5] and dielectric nitrides and oxides[6]. Meanwhile, significant progresses have been made on intrinsically non-layered three-dimensional (3D) materials (i.e. non-VDW materials) by synthesizing their 2D counterparts on selective substrates. Typical recent examples include: single atomic layer FeSe on Nb:SrTiO$_3$ substrate[7] boosting the superconductor T$_c$ via pronounced interface effect; III-V compounds[8], oxides[9], and PbTe/CdTe[10] 2D electron gas (2DEG) systems marked by their quantum Hall transport phenomena.

For the synthesis of the 2D form of naturally layered materials, several sophisticated methods have been developed including mechanical exfoliation, vapor deposition on catalytic substrates and van der Waals epitaxy[11-13]. However, for non-VDW 3D materials, the conventional chemical epitaxy by means of Molecular Beam Epitaxy[14] or Metal Organic Chemical Vapor Deposition (MOCVD)[15] is still the dominant strategy to obtain their 2D form. Such technique places stringent requirements[16, 17] on the choices of film-substrate combination in terms of lattice mismatch argument. This in turn narrows down the choices of materials for exploring innovative 2D physics. Fortunately, the discovery of 2D VDW epitaxy[13] initially for layered materials and later[18] for non-VDW 3D materials may open up a new and promising window for developing novel 2D materials or heterostructures (combined with the post transfer technique) free of substrate restriction. Progresses have been made on the growth of thin films of PbSe[19], Te[20], GaAs[21] and GaN[22] as well as vertical aligned ZnO nanorods that overcomes completely the fettering of the lattice mismatch[23, 24]. Nevertheless, most resultant film thickness (> ~ 20 nm) falls beyond a 2D criterion that is mandatory for enabling unique 2D physics[25]. The underlying reason leading to such observations is believed to be the non VDW bonding nature of the 3D non-layered materials, which favors the growth of island morphology rather than 2D layers with higher free energies. Ultimately, a couple of fundamental questions arise: (1) is it theoretically possible to grow large scale single or a few atomic layers films out of 3D non-layered materials by VDW epitaxy; and (2) what types of 3D materials are more feasible for the 2D growth?

In this report, we present our understanding on the 2D VDW growth of a 3D parent material methylammonium lead chloride ($MAPbCl_3$). $MAPbCl_3$ is a non-layered semiconducting material recently identified to be extremely efficient and promising for optoelectronics applications. Ultrathin (sub-10 nm) and large scale (a few tens of micrometers in lateral dimension) single crystalline 2D perovskite thin films on layered muscovite mica, a substrate very favorable for similar VDW growth for both layered and bulk materials[26, 27], were successfully synthesized by VDW epitaxy. Classical nucleation and growth model[28, 29] explaining conventional epitaxy has been modified to interpret the unique large-scale ultrathin 2D results in perovskite under VDW mechanism. The generalization of our VDW nucleation and growth model shows that the weak VDW interaction plays the key role in favoring the large-scale single crystalline 2D growth, which is typically retarded by a strong film-substrate interaction in the conventional epitaxy case. In addition, it demonstrates that ionic and metallic crystals with delocalized bonding characters have more tendency to form ultrathin structures compared with covalent materials with strongly localized bonds. Finally, with static and kinetic Monte Carlo simulations, we show that the fractal 2D morphology in perovskite materials precisely manifests the kinetic competition between VDW diffusivity and thermodynamic driving force. This unique phenomenon to VDW growth suggests a fundamental limit on the morphology stability of the 2D form of a 3D material. Our findings shed light on the growth strategies for the development of truly 2D structures out of their non-VDW 3D counterparts.

2. **Experiments and Results**

The $MAPbCl_3$ perovskite thin film was synthesized via a chemical vapor deposition (CVD) method with dual precursor sources. A schematic drawing and further details of the experimental setup can be found in supporting information Fig S1. Fig 1(a) shows the optical image of the well-aligned square perovskite sheets with high coverage over large area on mica substrate, with the typical results grown at a substrate temperature of around 220 °C and 9-10 cm away from the precursor. All crystal sheets align along one crystallographic orientation of mica. A closer examination (Fig 1(b)) reveals sharp perpendicular edges, consistent with the perovskite's cubic structure indicating high degree of crystallinity. Meanwhile, the uniformity in color suggests a smooth surface, confirmed by atomic force microscopy (AFM) images (Fig. 1c). Further observation of one of the largest square sheets (Fig 1(c)) shows a lateral dimension of a few tens of μm × a few tens of μm and thickness of sub-10 nm range. The thickness of the ultrathin films as determined by AFM cross sectional profiling (Inset of Fig. 1(c)) is down to 8.7 nm. Sheets with more 'whitish' color have a larger thickness ranging from 30 to 50 nm, but preserving the same smoothness (see supporting information Fig.S2a and b).Additional SEM images confirm the square-like morphology (see supporting information Fig. S3). The 2D sheets display different colors as the sheet thickness continues to increase (Fig 1(d)). Most sheets exhibit uniform color indicating good uniformity of the sheet thickness. Inset of Fig. 1(d) shows two typical thicker sheets of ~100 nm in thickness.

Interestingly, morphologies other than the square are also found popular. Fig 1(e) shows the typical "stair-like" zig-zag edges observed at almost each large sheet approximately 1-2 cm further away from the square ones. Such type of sheets featured by the fractal morphology are always found to be larger in size and could be grown to as large as several hundred microns scale (see supporting information Fig. S4a and b). Further, the classical dendritic morphology has also been observed (Fig. 1(f)) on the substrate farthest from the precursor (~13-14 cm away). It should be noted that here all dendritic branches share only two perpendicular directions. The versatility (square, zig-zag fractal, and dendritic) of the morphology and high reproducibility (~ 100%) of the synthesized perovskite by CVD VDW approach provide rich evidences on studying their growth thermodynamics and kinetics.

To confirm the perovskite structure and reveal its epitaxial relation with the mica substrate, transmission electron microscopy (TEM) and selective-area electron diffraction (SAED) characterization were performed. Considering the vulnerable nature of $MAPbCl_3$ to possible ion irradiation damage during typical TEM sample preparation processing by ion milling, moisture and organic solvents, we have developed a new TEM sample preparation method for the perovskite halide (Fig. 2(a)). As shown in Fig 2(a), the mica substrate with perovskite sheets growing on top was first peeled off by a few mica surface layers carrying the perovskite sheets with the aid of a polydimethylsiloxane (PDMS) stamp. Perovskite sheets/mica surface layers-covered PDMS was then pressed onto $LaAlO_3$ (LAO) substrate, upon which the peeled-off layers can be transferred. The relatively weaker interaction between perovskite sheets/mica surface layers and LAO substrate compared to the VDW force within mica layers made it more manageably and easily to transfer the perovskites sheets/mica surface layers (Occasionally only the perovskite sheets were transferred. See supporting information Fig. S5a and b) onto the copper grid simply by pressing the grid tightly on the LAO substrate.

Figs. 2(b) and (c) shows optical images of the as-transferred large-area "stair-like" fractal perovskite sheets (green samples in Fig. 2(b)) as well as square ones (e.g., the blue sample highlighted by the dash circle) on the copper grid. Fig. 2(d) shows a low-magnification TEM image of the square sheet marked in Fig 2(c). Inset of Fig. 2(d) shows its magnified optical image. Figs. 2(e) and (f) show the SAED patterns of both mica and perovskite sheets respectively, with the perovskite pattern taken from Fig S5a, and the [001] zone axis was identified for both samples. Despite being a monoclinic structure[30, 31], mica has a quasi-hexagonal lattice composed of potassium ions and aluminosilicates, as shown by the SAED pattern in Fig. 2(e). The perovskite sheet presents itself with a perfect set of cubic pattern, which agrees well with the nature of its cubic structure. Further electron diffraction indexing yields a lattice spacing of 5.69Å and energy-dispersive X-ray diffraction (EDX) spectrum (see supporting information Fig. S6) revealed the elemental signatures of N, Cl and Pb, both of which confirmed the perovskite phase of the $MAPbCl_3$[32]. A SAED pattern as shown in Fig 2(g) taken from a perovskite sheet/mica surface layers (spot (g) in Fig. 2(d)) illustrated the epitaxial relation between mica and the perovskite halide featured by a combination of two sets of diffraction patterns (square and hexagonal). Combined with our optical evidences in Fig. 1 and further SAED characterizations (see supporting information Fig. S7a and b), we have revealed a consistent and universal VDW epitaxial relation between mica and $MAPbCl_3$ – mica (001) ∥ perovskite (001), mica (200) ∥ perovskite (200) (5º offset) and mica (020) ∥ perovskite (010) (5º offset) - for all three perovskite morphologies (square, "stair-like" fractal, dendritic). The 5º offset along the two in-plane directions may result from the different lattice types of the two materials and the need for the two to maximize the VDW bonding by choosing the optimized orientation[33]. It can be concluded from the indexing of Fig. 2(f) and (g) that the two perpendicular free facets of perovskites are (100) and (010), respectively, which represent the thermodynamic preferable configuration of the perovskite halide. To better understand and illustrate the VDW epitaxial relation between mica and the perovskite halide, we present a tentative atomistic model[34], as shown in Figs. 2(h)-(j), where the perspective, normal and side views of the perovskite/mica heterostructure are presented, respectively.

The VDW growth nature of perovskites on mica is further supported by several other evidences and arguments: (1) the ***layered nature*** of mica barely supports chemical epitaxial growth; (2) if the chemical bond does exist between perovskites and mica, it is more natural for the perovskite to be grown on mica with perovskite {111} surface oriented since it is hexagonal and should fit better with the substrate; (3) another strong evidence that may rule out chemical epitaxy is the absence of misfit dislocations for thick films, a most common way to relax the strain created by the substrate and film; (4) with the chemical epitaxy excluded and the epitaxial relation still observed, based on the layered nature of mica, VDW epitaxy remains a very feasible explanation; (5) the fact that we could accidentally peel off perovskite from mica also indicates the existence of a weak VDW force between mica and perovskites.

### 3. Discussions and Modelling

To understand the growth of the ultrathin non-VDW perovskite sheets on mica by VDW mechanism, a quantitative VDW nucleation and growth model is proposed. The most abrupt difference between conventional epitaxy and VDW epitaxy is the chemical activity of the substrate surface – conventional epitaxy involves highly active substrate surface while VDW substrate is almost inert. In a nucleation process, adatoms on the substrate surface bond with both the substrate characterized by adsorption energy, $E_{ad}$, and other adatoms characterized by interatomic bonding energy, $E_i$, to obtain sufficient energy so as to form a nucleus that can continue to grow. Fig 3(a) schematically describes the atomistic nucleation process in a conventional and a VDW epitaxy process. In both cases the nutrient atoms get adsorbed and diffuse on the substrate surface. The strong chemical bond between adatom and substrate in the conventional case leads to a quite small critical nucleus (down to single atom); the deep bonding state and the high anti-bonding position in this case result in a large diffusion barrier ($E_d$). While for the VDW case, the inert substrate is free from dangling bonds and only small dipole movements exist. Thus $E_{ad}$ is typically characterized by the weak VDW energy. Consequently, the weak VDW $E_{ad}$ leads to a much harder nucleation process since more "simultaneous" collisions of adatoms on the substrate are needed. In addition, the higher energy

VDW bonding state makes diffusion barrier extremely low, which is indeed confirmed by recent simulation work in the GaAs-graphene material system[21].

For a quantitative understanding on the VDW process, we present a modified nucleation and growth model. In a nucleation process, adatoms with $E_{ad}$ will have a lifetime on the substrate as[28]:

$$\tau_a = v^{-1}\exp(E_{ad}/kT) \qquad (1)$$

where $v$ is the vibration frequency and $k$ the Boltzmann constant. Multiplied with the incoming atom deposition rate, $R$, it gives the concentration of single adatom on the substrate, $N_1 = R\tau_a$. Based on Walton's model of thin film growth in conventional epitaxy[35], at low coverage, the concentration of the cluster with $i$ atoms, $n_i$, can be related with $N_1$ and the interatomic bonding energy $E_i$ by:

$$n_i = \frac{N_i}{N_0} = \left(\frac{N_1}{N_0}\right)^i \exp\left(\frac{E_i}{kT}\right) = \left(\frac{R\tau_a}{N_0}\right)^i \exp\left(\frac{E_i}{kT}\right) \qquad (2)$$

where $N_0$ is the density of adsorptions sites available on the substrate. The formula contains a statistical term, $\left(\frac{R\tau_a}{N_0}\right)^i$, that relates the probability of finding $i$ atoms holding together, and a thermodynamics term, $\exp\left(\frac{E_i}{kT}\right)$, that depicts the energy benefit of the cluster, both of which make the model intuitively sound. On the other hand, classic nucleation theory[36] gives a similar form of $n_i$ in terms of $N_1$ and the Gibbs free energy change, $\Delta G_i$, when a cluster of $i$ atoms is formed:

$$n_i = \frac{R\tau_a}{N_0} \exp\left(\frac{-\Delta G_i}{kT}\right) \qquad (3)$$

The free energy term takes into account the energy benefit from vapor into solid as well as the energy penalty resulted from the creation of free surfaces in the shape of either a cap or a disk. Therefore the cluster with the highest $\Delta G_i$ would be the most unstable one and thus the critical nucleus size $i^*$, i.e., upon capturing a new atom, the nucleation barrier $\Delta G_i^*$ would be overcome and a stable nucleus formed. Lewis has proved earlier[37] that the two models – Walton's and the classic one - are equivalent with the former being more suitable at small number of $i$ since it is hard to really form a free surface with just very few atoms. Therefore, mathematically one could get $\Delta G_i^*$ and the critical nucleus size using Walton's model, i.e., the cluster with the lowest coverage.

In the case of the perovskite halide, from the enthalpy of sublimation data of $PbCl_2$ and MACl, the bond energy of Pb-Cl and MA-Cl are estimated to be 0.515 eV and 0.298 eV, respectively (supporting information Supplementary Text 1). Therefore, $E_i$ could be acquired by maximizing the energy benefit at a given cluster size. Fig. 3(b) plots how the coverage of different size of clusters (1 - 7 illustrated on the right side of Fig. 3(b)) differs under different adsorption energy at a typical atomic vibration frequency $10^{12}\,s^{-1}$, a moderate deposition rate R = 1 monolayer (ML)/min and T = 500 K (the temperature of the substrate where most perovskite sheets are found). For simplicity concern, we assume the same $E_{ad}$ for all three kinds of adatoms on the surface. The ball-and-stick models on the right of Fig. 3(b) show the optimal configuration of every cluster. According to Fig. 3(b), the coverage decreases exponentially as $E_{ad}$ decreases as a general trend for all $i$. At a given $E_{ad}$, the cluster with the lowest coverage characterizes the critical size. When $E_{ad} > 1$, corresponding to a conventional epitaxy case, $i^* = 1$ and nucleation proceeds easily (with a coverage value above $10^{-5}$. At low $E_{ad}$ value such as 0.5 eV corresponding to a typical VDW adsorption[38], we obtain a critical nucleus size of 6 and a coverage of ~ $10^{-25}$, indicating almost impossible nucleation event (adsorption site density is ~ $10^{19}/m^2$). Clearly, the low VDW interaction allows a larger kinetic window (or dynamic range) for the possible growth of large scale single crystalline 2D sheet due to its extremely difficult nucleation event. Fig 3(c) plots the change of $\Delta G_i$ with $i$ framed in a classic model converted from Walton's model in Fig. 3(b). Two $E_{ad}$ values of 0.77 and 0.63 eV are exemplified to further illustrate the identification process of $i^*$.

Further, the explicit nucleation rate, $U$, is determined. The rate is proportional to the density of the critical nuclei, $N_{i^*}$, which is one atom away from a stable one, and also the number of "free" single adatoms, $N_1$. Therefore $U$ is given by[28]:

$$U = \frac{dN}{dt} = DN_1 N_{i^*} = DR\tau N_{i^*} \qquad (4)$$

where $D = \frac{1}{4}a^2 v \exp(-E_d/kT)$ is the surface diffusivity of adatom on the substrate. The lifetime $\tau$ could be either limited by the re-evaporation of adatoms, which makes $\tau$ follow the same form of eq.(1) when the density of stable nuclei is small or limited by being captured by stable nuclei before adatoms re-evaporate, which leads the lifetime to be[39] $\tau_c = 1/DN$, where $N$ is the overall density of nuclei. Integration of eq.(4) yields the density of nuclei after a certain deposition time $t$:

$$N = DR\tau_a N_{i^*} t \quad \text{(re-evaporation rate-limited)} \tag{5}$$

$$N = \sqrt{2RN_{i^*}t} \quad \text{(nuclei capturing rate-limited)} \tag{6}$$

Nucleation rates under both scenarios are plotted against $E_a$ in Fig 3(d) with $t = 1s$. It is clear in both cases the nuclei density experiences an exponential decrease with $E_{ad}$, with the additional drop in slope (log scale) resulting from the change in the critical nucleus size, as marked by different colors on the graph. The quantitative results above clearly show how the change in adsorption energy dramatically affects the nucleation process, which is the fundamental reason why VDW epitaxy is a unique process differing from conventional epitaxy. Notably at around $E_{ad}$ = 0.5 eV (the approximate value of a single bond), a rough estimate of the adsorption energy in VDW case[38], only one nucleus may be expected to exist in a reasonably large area of 0.2 cm². This is indeed confirmed in our experiment. As shown in Fig. 3(e), despite being separate square sheets, the fact that their parallel edges do not follow the 6-fold symmetry of mica is a strong indication that all of them started with only one nucleus; whereas by increasing the supersaturation via fast substrate cooling in which case from eq.(1) $\tau_a$ increases exponentially and so does the single adatom concentration, we obtained square sheets with obvious different orientations, suggesting multiple nuclei, as being marked by coordinates with different colors in Fig. 3(f). The way to determine whether it is single nucleus or multiple nuclei is as follows: statistically, for multiple nuclei, we would not observe the growth in Fig. 3(e) since all 6-fold orientations of mica have equal probability for nucleation and the square films should align along several different orientations rather than only one orientation; similarly, epitaxial but different orientations of square films in Fig. 3(f) rules out the possibility of single nucleus. The same phenomenon can be found in the dendritic morphology (see supporting information Fig. S8a and b). Despite this, the epitaxial relation found in Fig. 2 still persists in Fig. 3(f). Fig. 3(g) illustrates how the 6-fold symmetry of mica and the 4-fold symmetry of perovskite can account for the seemingly-disordered but indeed epitaxial growths in Fig. 3(f), where the 5° offset is taken into account. In other words, while preserving the 5° offset (which can take place in two directions as shown in the arrows in Fig 3(f)), the perovskite nucleus can be oriented in 6 equivalent directions corresponding to the symmetry of pseudo hexagonal mica, thus giving rise to the different possible orientations among different nuclei when nucleation process is encouraged. To further confirm our analysis and compare the conventional epitaxy and VDW one, we conducted perovskites growth on different types of substrates including silicon with very active surface, graphene on silicon, graphene on mica, and non-VDW material -seeded mica (see Fig S9a-f and the corresponding discussion in supporting information). $PbCl_2$ with stronger chemical bonds in average than the perovskite halide was also grown on mica (see Fig S13f and the corresponding discussion in supporting information). The observed results show great agreement with our kinetic analysis. Meanwhile, it is also noticeable that $PbCl_2$ is formed with a complete different morphology and at a different location (4 - 5 cm away from the furnace) compared with the square sheets. This observation helps to rule out any possibility of non-uniformity in the composition of the as-grown perovskite thin film such as incorporation of $PbCl_2$.

So far the analysis has been limited to the kinetic process of the first layer, but what parameters dictate the evolution of 2D films along vertical direction? Assuming films grow along vertical directions via layer-by-layer growth mechanism, the second layer growth would require a new round of nucleation which is again affected by the adsorption energy, only on a different "substrate" – first layer. As schematically shown in Fig 3(h), for intrinsically-layered VDW materials, due to their weakest VDW-type adsorption energy, the nucleation process of the second layer would be as strenuous as the first layer. Adsorption energy of 3D material, on the other hand, would recover to their "normal" value as the film thickness increases. The recovery process would be very fast for localized-bonds materials (e.g. covalent) that do not have significant long-range interaction[40] but slower for the delocalized-bonds materials (e.g. ionic and metallic materials) in which long-range interactions are substantially important[41, 42]. The relatively smaller adsorption energy of the thinner delocalized-bonds materials would help suppress the nucleation rate along vertical direction. Therefore, thinner structures of delocalized-bonds materials are more likely to promote the 2D growth.

To reveal the feasibility of atomically thin 2D film via VDW epitaxial growth, we further quantify the ratio of the nucleation rate of the second layer, $u_n$, and the lateral growth speed, $v_l$, since monolayer film can be obtained as long as $u_n$ is curbed. The mean free path of an adatom on the surface, i.e., the magnitude of the random walk it experiences before re-evaporation, is given by $L = \sqrt{D\tau_a} = 0.5a_0 \cdot \exp((E_{ad} - E_d)/2kT)$ ($a_0$ is the single hop distance of adatom on substrate). For simplicity, we assume the first layer to be a disc with radius $r$ and therefore the area where adatoms could potentially incorporate into the lattice will be the ring region around the disc with a width of $L$, the dark yellow region marked in Fig. 3(i). The nucleation event may occur on the already-formed disc. With further derivation (supporting information Supplementary Text 2), the lateral growth rate is given by:

$$v_l = dr/dt = \frac{RL}{2N_0}(\frac{1}{2} + \frac{1}{6}\frac{L}{r}) \tag{7}$$

The nucleation rate follows the same analysis as previous one but with a different adsorption energy, $E_{ad2}$, and the area of the disc is used to yield the exact number of nuclei instead of the density value. Fig 3(j) shows the 3D surface plots of how $\ln(u_n/v_l)$ varies with the deposition rate $R$ and the size of the disc $r$ ($u_n$ and $v_l$ are normalized to be the same dimension here) for three types of materials. $E_{ad2}$ values of 1.5, 0.7 and 0.4 eV are used, which to some extent represent the rigid covalent bond, weak delocalized ionic bond and VDW bond, respectively. In general a lower $E_{ad2}$ retards the nucleation process and thus a smaller $u_n/v_l$ ratio is obtained, making it more feasible to grow monolayer film. For all three cases, at a fixed size, $v_l$ has a linear dependence on the deposition rate, while $u_n$ is proportional to $R^{i^*+1}$. Therefore mathematically at sufficiently low $R$, $\frac{u_n}{v_l} \ll 1$ would always hold and monolayer thickness could be always achieved. Such relation is explicitly shown in Fig 3(k) with $r = 50$ nm. Therefore, according to above analysis, among all non-VDW materials systems discussed, weak delocalized-bonded 3D materials would mostly favor the 2D growth via the VDW mechanism. On the other hand, it is hard, but not impossible, to obtain 2D growth for materials with higher cohesive energy since nucleation on the second layer tends to take place very easily—but still can be curbed as long as supersaturation is well-controlled.

Based on above analysis, for the perovskite halide with extremely low cohesive energy and strong ionic character, a significant percentage of large scale monolayer growth with proper growth conditions is expected. However experimentally, the thinnest films we could find are always around 8 nm. To understand this, further analysis assisted with static and kinetic Monte Carlo simulations on the growth kinetics of fractal morphology is conducted.

Figs. 4(a)-(e) present five types of typical growth morphologies of the perovskite halide

obtained at different locations of substrates during several experimental attempts, with the scale bar adjusted to render a better visualization of the morphology transformation. Fig. 4(a) shows the square morphology with four side facets being {100}, which follows Wulff construction rule at the 2D space. For all other cases in Fig. 4(b)-(e), the film propagates along the <110> direction in a fractal manner rather than expanding along <100> direction by following the surface energy argument. Apparently, the fractal morphologies are a consequence of kinetics. The rate competition is proposed to stem from adatoms' VDW-type diffusion and perovskite films' anisotropic surface free energy. We attribute the <110> fractal growth direction to be a consequence of the anisotropy in the adatom-capturing capability of different sites at the film edges. At the initial stage of film growth, surface energy dominates and a square shape should be expected since it is easy to achieve local equilibrium. With a square shape, the region where adatoms could potentially incorporate into the matrix is marked by both the yellow and blue colors in Fig. 4(f). The outside border of the blue and yellow regions is formulated by the diffusion length ($L$) of the adatoms. According to random walk theory[43], the probability that the adatom terminates at a certain vector $\vec{r}$ from the original point after random walk scales with $\sim \exp(-|\vec{r}|^2/4Dt)$, therefore the edge site that is closest to the deposited atom has the highest probability to capture it. Consequently, the corner sites of the square film would have higher probability to be incorporated by the adatom. A static Monte Carlo simulation was carried out to prove our hypothesis in which atoms, randomly deposited around a square film with the side-length $a$, conduct random walk until they are captured by the matrix or a certain number of jumps is reached. Fig 4(g) plots the simulation results on capturing probability (i.e. the number of atoms captured at one site divided by the total number of atoms captured) at different sites under different $L/a$ ratios. It can be seen that at small $L/a$ ratio (e.g. 0.1) a relative uniform capturing probability distribution along the corner vertex to the middle of the square side is obtained, which is a result of the minimized area ratio of blue region over yellow one. For a larger $L/a$ value (e.g. 1), higher capturing probability at the corner is observed. In this case, our analysis is mainly based on the assumption that surface diffusion is much faster than crystal edge diffusion. Such assumption requires a considerable size of diffusion path "$a$". It should be noted that when $a$ is extremely small, i.e., at the initial stage of the growth, the Wulff ripening process would dominate during crystal growth as diffusions along the crystal edges become much easier. This would lead to a thermodynamic stable structure rather than the kinetic fractal morphology. This also explains why the fractal morphologies are always of a larger size. A more visualized illustration of the anisotropy in the capturing rate of different edge sites is shown as a 3D chart in Fig 4(h), with the inset being the normal view. Further, we simulated the adatom capturing rate of different edge sites of a fractal morphology in Fig. 4(b), as shown in the 3D chart of Fig 4(i). Clearly, all corner vertices have higher capturing rates than non-vertex sites; inner corner sites have slightly lower capturing rates than outmost corners. Higher capturing rates indicate faster growth. Therefore, the corner preference serves as the morphology instability source that induces fractal growth. Meanwhile, local equilibrium favors the formation of {100} facets in perovskite film and therefore often fractal growth morphologies with well-faceted local structures (i.e. "stair-like" zig-zag contours) are observed. The large-scale local equilibrium occurs more easily in VDW epitaxy process as the VDW-type diffusivity is orders higher in magnitude than the one in conventional epitaxy. To further support our argument, kinetic Monte Carlo simulation is conducted that precisely predicts the well-faceted fractal morphology by including both anisotropic capturing and equilibrating processes (see supplementary video 1 and 2). Such fractal growth over large area is different from earlier observation of angstrom scale fractal growth of metal where diffusion is extremely limited[44, 45], but often found in the formation of snowflakes, where the diffusivity of water molecules is high as well[46].

The high VDW-type diffusivity (surface diffusivity and edge diffusivity) and low cohesive energy of perovskites are further supported by more evidences. Fig 4(j) and (k) shows the morphology change of the same sample across four weeks at room temperature and ambient atmosphere. By aging, thinner films at location 5 in Fig. 4(j) sacrifice themselves (5' in Fig. 4(k)) and participate in the construction of the fractal structure of thicker films (1 – 4 in Fig. 4(j) and 1' – 4' in Fig. (k)). Further, a few days later the small fractal structure at 4' transformed to a square one, possibly due to the easiness to reach local equilibrium (see supporting information Fig. S10). Similar change was observed universally where a sharp discrepancy in thickness existed (see supporting information Fig S11a - d). A live video was also recorded to show rapid real-time change of morphology overnight at room temperature (see supplementary video 3). Accordingly, uniform and separate square films (Fig. 1(a) and (b), Fig. 3(e) and Fig.

4(i)) with identical crystallographic orientation could be regarded as a consequence of the ultrahigh VDW diffusivity during materials growth where substrate temperature is high (220 ºC). In other words, the sacrificing of thinner films and construction of thicker films proceed much faster during material growth, which could precisely explain why many crystals share a single nucleus and these crystalline sheets line up along <110> direction (blue arrows in Fig. 4(i) and (m)). More evidence on the manipulation of such "relaxation" process could be found in supporting information Fig S–12,13 and Fig S14a - e). All these evidences illustrate that the absence of mono- or a few-layer 2D results (i.e. << 8 nm) is very likely caused by the high VDW-type diffusivity and low cohesive energy of perovskites, which lead to film thickening during and after materials growth. Therefore, for VDW monolayer growth of perovskite film, i.e. to prevent the film thickening process, further strategies are required. Examples are: (1) coating of second dielectric layer on top of perovskite layer, and (2) reducing growth and sample storing temperature. Indeed, first strategy may lead to atomic artificial structure in a great analogy to layered superconducting cuperates[47], irridates[48] and nickelates[49]. In fact, these two strategies have been found useful in preparing monolayer oxides films either in wet chemistry or vacuum deposition growth.

4. Conclusion

In summary, we show the successful growth of sub-10 nm perovskite halides on mica by VDW epitaxy mechanism. Our proposed VDW nucleation and growth theory precisely explains that it is the weak VDW film-substrate interaction and low cohesive energy of the perovskite halide that lead to large scale single crystalline ultrathin 2D growth. The model also illustrates that ionic and metallic crystals with weak delocalized bonds tend to more favor 2D growth than strongly covalent materials. Finally, combined with Monte Carlo simulation, our model shows that the VDW-type diffusivity and low cohesive energy of perovskites determine the ultimate kinetic thickness of the perovskite films. Fractal morphology is a result of the competition between thermodynamic driving force and VDW diffusivity. Our findings reveal the possibility to grow mono- and a few-layer 2D film via VDW epitaxy out of non-VDW 3D materials[9, 50].

**Supporting Information**

Experimental setup, methods, materials, optical analysis, EDX and Monte Carlo simulation and real-time morphology change videos can be found in the supplementary information.

**Acknowledgements**

J. S. and Y. W. were supported by J.S.'s start-up fund from Rensselaer Polytechnic Institute and National Science Foundation under grant CMMI 1550941.

**Figures and Figure Captions:**

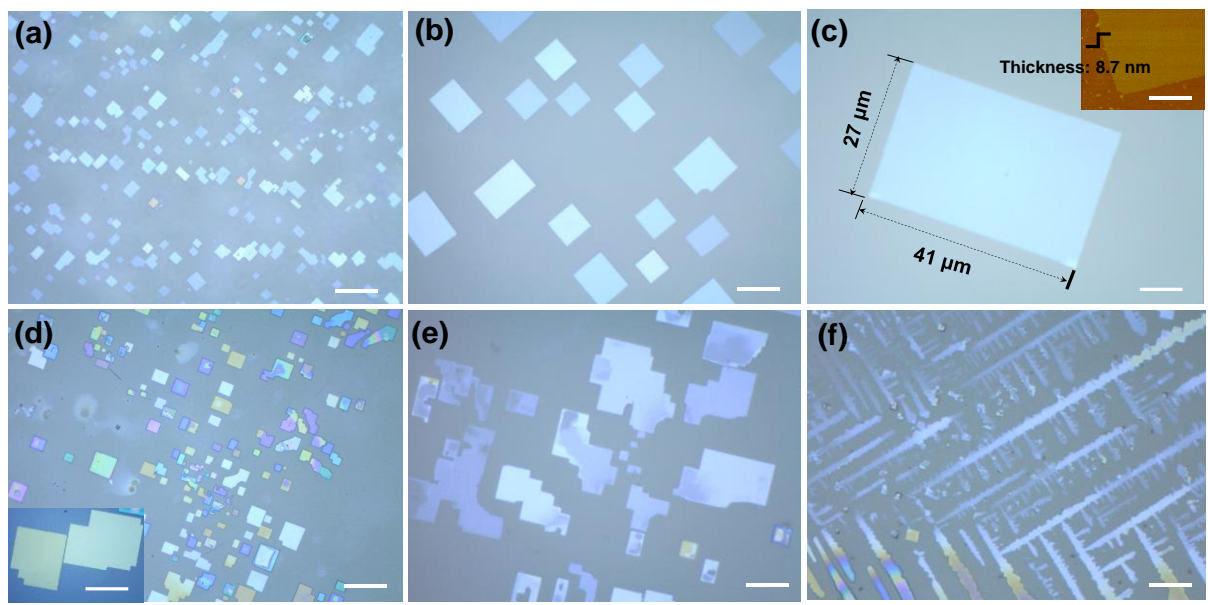

**Figure 1. Morphology of as-grown perovskite (PVK) thin films.** (a)-(b) Optical microscope images of well-aligned square sheets covering large areas of mica substrate; (c) An optical image of an individual large-scale ultrathin sheet (inset showing the atomic force microscopy measurement of the film thickness); (d) High coverage colorful square sheets with larger thickness (inset showing the magnified view of two typical examples); (e) "Stair-like" zig-zag morphology of perovskite sheets; (f) Dendritic morphology expanding in two perpendicular directions. Note none of the images are false colored (Scale bar: (a) 50 μm (b) 10μm (c) 10 μm (d) 50 μm (inset: 20 μm) (e) 10 μm (f) 50 μm).

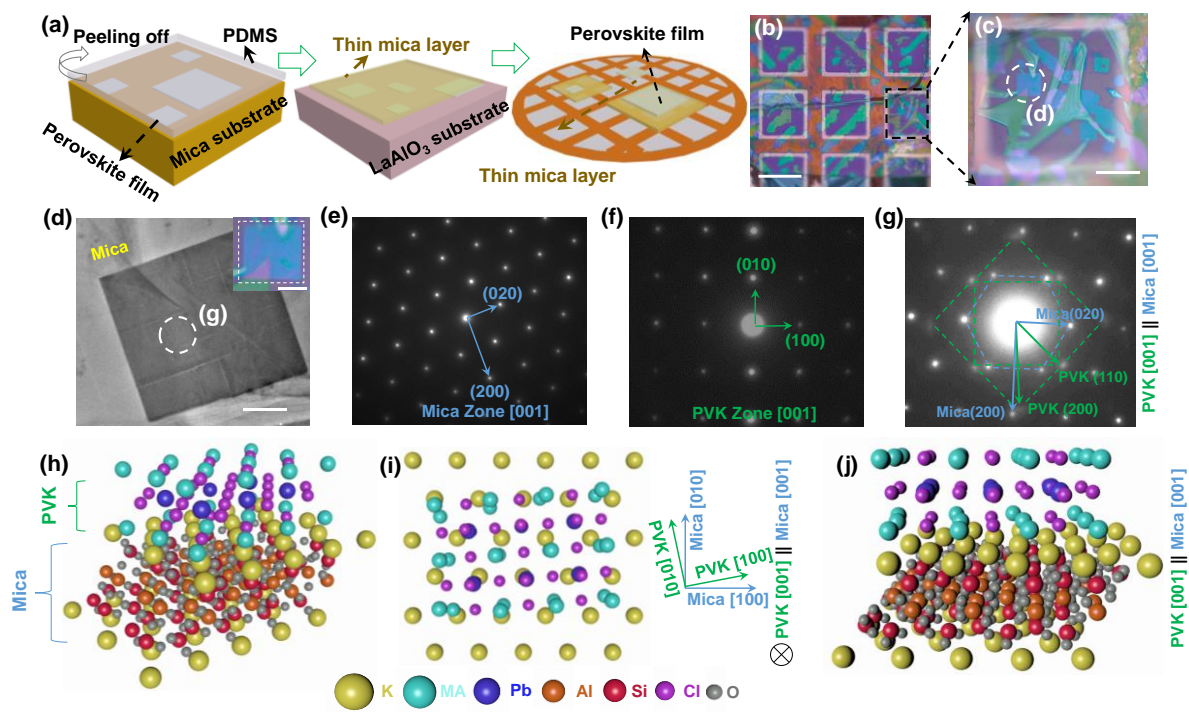

**Figure 2. Structural analysis of perovskite (PVK) thin films growing on mica.** (a) Schematic drawings of the "peeling-off" TEM sample preparation technique of as-grown PVK thin films; (b) An optical microscope image of PVK/mica sample on the copper grid; (c) A magnified optical image of the area marked by dashed box in (b); (d) A low magnification TEM image of a square sheet marked in (c), with the inset showing the magnified optical image; (e)-(f) Electron diffraction patterns of separate mica (hexagonal) and PVK (cubic) with zone axis both as [001], (f) was taken from Fig S5a; (g) Electron diffraction pattern taken at the dashed circle in (d) containing both mica and PVK lattices revealing the epitaxial relation, where an approximate 5 ° offset in the mica/PVK (200) direction is present; (h)-(j) Atomistic model of mica/PVK epitaxy from a perspective (h), normal (i) and side (j) view. Legends at the bottom mark the atom species. (Scale bar: (b) 50 μm (c) 10 μm (d) 1 μm (inset: 2 μm))

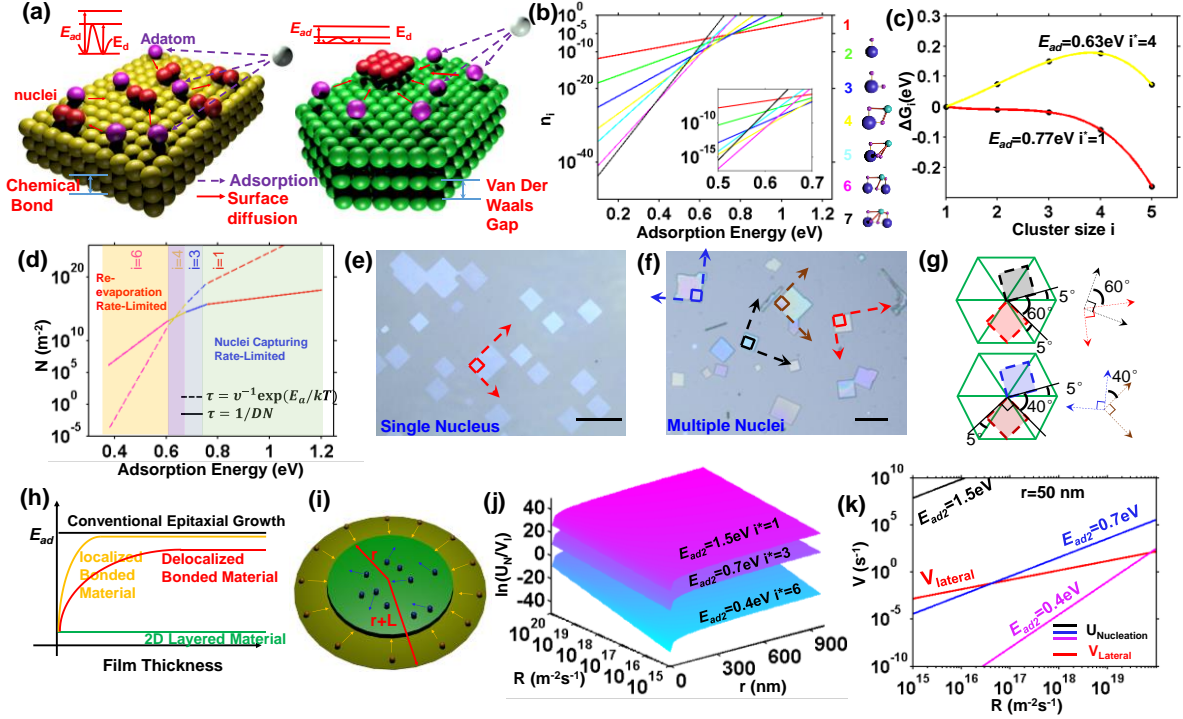

**Figure 3. Van der Waals epitaxial nucleation and growth model of perovskite thin film.** (a) Schematic drawing that shows different atomistic processes on a conventional (left) and VDW (right) substrate in terms of adsorption energy ($E_{ad}$), diffusion barrier ($E_d$) and nuclei sizes; (b) The plot shows how coverage of a cluster with a certain size $i$ changes with $E_{ad}$ (Inset shows a magnified version of $E_{ad}$ at 0.5-0.7 eV). Each color line reflects a different cluster size, as shown by the corresponding numbers on the right. The ball-and-stick models demonstrate the optimized configuration of each cluster, with the atom legends same as Fig (2); (c) Plot of $\Delta G_i$ vs. $i$ at $E_{ad}$ = 0.77, 0.63 eV, showing the critical size, $i^*$, identification process; (d) Plot of nuclei density versus $E_{ad}$ under capturing rate-limited (solid line) and re-evaporation rate-limited (dashed line) assumptions. Change of $i^*$ is shown by different color lines corresponding to (b); (e)-(f) Optical images showing single nucleus (e) and multiple nuclei (f), with color arrows indicating different aligning orientations; (g) Illustration that the multiple nuclei in (f) hold the same epitaxial relation observed in Fig (2) through the crystal symmetry and 5° offset; (h) Schematic drawing of how $E_{ad}$ changes with thickness for three types of materials; (i) Schematic drawing of the disc (green) model used to estimate the lateral growth rate and the yellow region outlining the adatoms-capturing area; (j) 3D surface plots showing how deposition rate ($R$) and disk size ($r$) determine the ratio of vertical nucleation ($u_n$) and lateral growth ($v_l$) with three typical $E_{ad2}$ values; and (k) plot of $v_l$ and $u_n$ versus $R$ at a fixed disc size ($r = 50$ nm) at the same $E_{ad2}$ values as in (j). (Scale bar: (e) 50 μm (f) 20 μm)

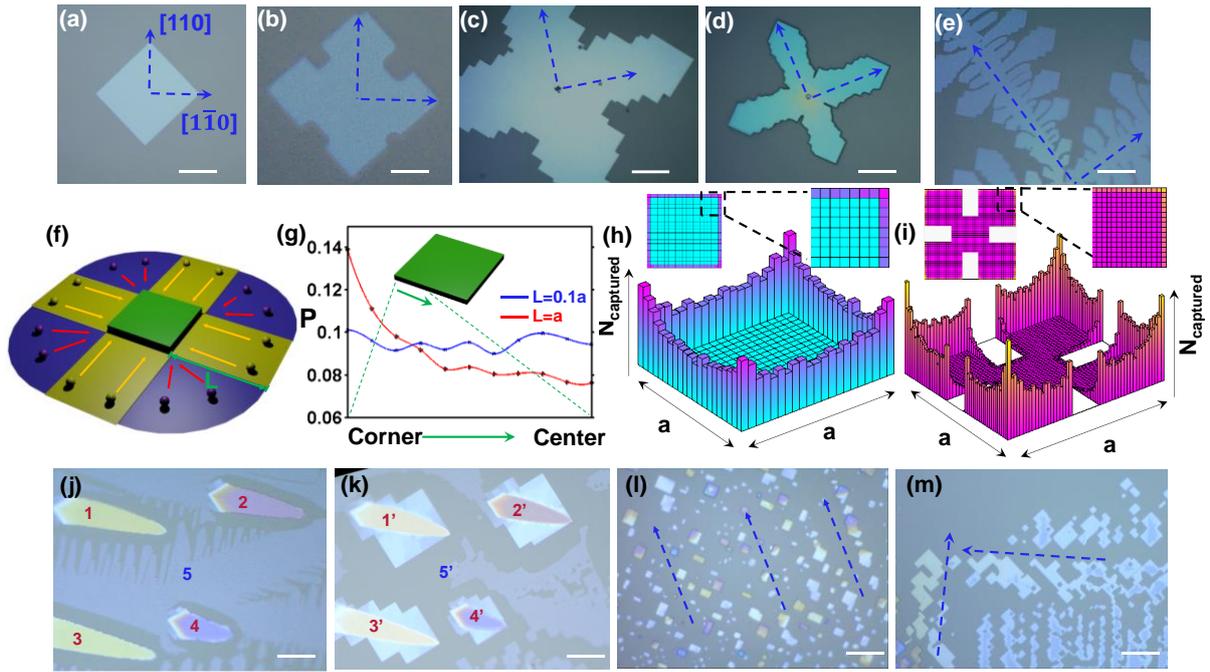

**Figure 4. Fractal evolution of perovskite thin films in Van der Waals epitaxy.** (a)-(e): morphology evolution of (a) a square sheet with {100} facets through (b)-(d) protrusion along the <110> direction to (e) fractal dendrite; (f) Schematic drawing hypothesizing that the corner vertices of a square sheet have higher adatoms-capturing rate than the inner region of the square edges; (g) Plot of capturing probability ($P$) versus edge sites from corner to center of a square perovskite seed with Monte Carlo simulation under $L = 0.1a$ (blue line) and $L = a$ (red line); (h)-(i) 3D bar chart showing Monte Carlo simulation result for a square (h) and fractal (i) seed under $L = 0.1a$. Insets: (magnified) normal view of the 3D bar chart; (j)-(k) Morphology evolution of perovskite films at room temperature within four months. Samples at locations 1 - 4 evolved from (j) 2D cone-like structure to (k) "stair-like" zig-zag structure and film at location 5 disappeared; and (l)-(m) As-deposited perovskite square sheets (separated in (l) and semi-connected in (m)) showing obvious lining orientations marked by the blue dashed lines. (Scale bar: (a) 10 μm (b) 5 μm (c) 20 μm (d) 10 μm (e) 20 μm) (j)- (m) 50 μm).

**For Table of Contents Use Only**

**Title: Two-Dimensional Van der Waals Epitaxy Kinetics in a Three-Dimensional Perovskite Halide**

Authors: Yiping Wang[1], Yunfeng Shi[1], Guoqing Xin[2], Jie Lian[2], Jian Shi[1*]

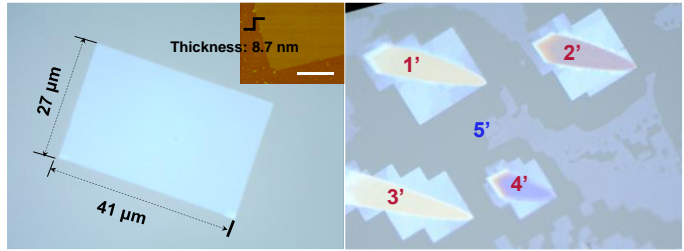

A brief synopsis: we report the two-dimensional van der Waals epitaxial growth of $CH_3NH_3PbCl_3$ perovskite thin films on mica. Large scale ultrathin perovskite films are proposed to be a unique consequence of van der Waals epitaxy.

# Supplementary Information

## Two-Dimensional Van der Waals Epitaxy Kinetics in a Three-Dimensional Perovskite Halide

Yiping Wang[1], Yunfeng Shi[1], Guoqing Xin[2], Jie Lian[2], Jian Shi[1*]

1. Department of Materials Science and Engineering, Rensselaer Polytechnic Institute, Troy, NY, 12180

2. Department of Mechanical, Aerospace, and Nuclear Engineering, Rensselaer Polytechnic Institute, Troy, NY, 12180

[*]Corresponding author: shij4@rpi.edu

**Contents:**
Material and Methods
Supplementary Text ST1,ST2
Supplementary Figures S1 – S14
Reference

# Material and Methods

## Chemical vapor deposition (CVD) synthesis of MAPbCl$_3$

Powdered Lead(II) chloride(PbCl$_2$, 99%, Sigma-Aldrich) was placed in the furnace heating center with the heating temperature controlled at 360-380° C, while MACl (99%,Merck KGaA), the second precursor, was placed about 6 cm away from PbCl$_2$ in the upper stream due to a lower melting point. Fresh cleavaged muscovite mica substrates (SPI Grade V-5) with (001) face exposed were placed in the downstream. Prior to deposition, the base pressure of the system was pumped to 0.3 Torr after which a 30 sccm of Argon was flowed to maintain the pressure at 120-160 Torr before deposition. The chamber temperature rose from room temperature to the deposition temperature rapidly in 5 min. The deposition process lasted for 20 minutes before finally the furnace was shut down. The perovskite(PVK) film was found, in most cases, present on the third substrate with a temperature at about 200° C. Schematic drawing of the experimental setup can be found in Fig S1.

## Characterization of Perovskite thin film

Morphology of perovskite thin film was characterized by a Nikon Eclipse Ti-S inverted optical microscope and JEOL JSM 6330F Field Emission Scanning Electron Microscope. Multimode$^{TM}$ Atomic Force Microscope is used to obtain the film thickness. Transmission Electron Microscope JEOL JEM-2010 is used to characterize the structural and epitaxial relation of the as-grown PVK thin film.

# Supplementary Text

## ST1 Estimation of the Pb-Cl and MA-Cl bond energy
Estimation of bond energies are based on the assumption that only the nearest neighbors are concerned. The bond energy is estimated from the enthalpy of sublimation (addition of enthalpy of fusion and vaporization) and we assume when vaporized, the material breaks all the bonds.

### *Pb-Cl Bond*
$PbCl_2$ has a enthalpy of fusion and vaporization of 122 and 27.5 kJ/mol according to the literature *(1)*. Structure analysis *(2)* of $PbCl_2$ shows a orthorhombic Bravais lattice and every $Pb^{2+}$ cation has three nearest neighbors at a distance of approximately 2.80-2.91 Å, which is very close to the Pb-Cl distance in $MAPbCl_3$ (2.85 Å). Therefore it is a good assumption to take into account 3 mol of effective Pb-Cl bonds broken per mol of $PbCl_2$ when sublimated. The analysis above gives $\varepsilon_{Pb-Cl} = 0.515\ eV$.

### *MA-Cl*
Enthalpy of fusion data for MACl is not available most probably because of its hydrophilic nature. It has an enthalpy of vaporization of 115 kJ/mol *(3)*. It has a tetragonal structure *(4)* where $Cl^-$ sits on all the vertices and two of the face centers and $MA^+$ sits in the rest face centers, leaving each kind of ion 12 nearest neighbors, among which 8 is the opposite kind. Here we roughly assume the repulsion of the same ions cancels out the attraction of a cation and an anion, therefore leaving a net value of 4 MA-Cl bonds. Also, in MACl crystal the distance between MA and Cl is found to be 3.93 Å, which is also close to their distance of 4.02Å in $MAPbCl_3$. The analysis above gives $\varepsilon_{MA-Cl} = 0.298\ eV$.

## ST2 Derivation of the lateral growth speed $v_l$
Under disc assumption, the effective deposition area, as discussed in the main text, is:
$$S = \pi(r+L)^2 - \pi r^2 = \pi(L^2 + 2rL) \qquad (1)$$
Assuming the number of atoms captured by the disc to be proportional to the number of atoms deposited in region S per unit time by a proportional constant $k$:
$$\frac{dN}{dt} = k\pi RL(L + 2r) \qquad (2)$$
To further determine the value of $k$, we further examine two extreme conditions where atoms are deposited either right beside the existing film or at the far edge of the effective deposition area. For the first one, atoms have approximate 0.5 probability to be captured while in the other case, the probability is almost 0. Since in our case $L$=25nm while $r$ could be as large as 1 μm, it is reasonable to assume that the probability drops linearly as atoms are deposited further away from the edge of the film. Therefore, by integration we obtain:
$$\frac{dN}{dt} = R\int_0^L 2\pi(r+x)dx\left[0.5 - 0.5\frac{x}{L}\right] = \pi RL\left(\frac{r}{2} + \frac{1}{6}L\right) \qquad (3)$$
which is very close in form to eq.(2) when $L<<r$, $k$=1/4, consistent with 2D random walk.
When atoms captured by the disc film are uniformly distributed, the contribution to the lateral growth would be:
$$dN = N_0 2\pi r dr \qquad (4)$$
Substituting into eq.(3), we have:
$$v_l = \frac{dr}{dt} = \frac{RL}{2N_0}\left(\frac{1}{2} + \frac{1}{6}\frac{L}{r}\right) \qquad (5)$$

# Supplementary Figures

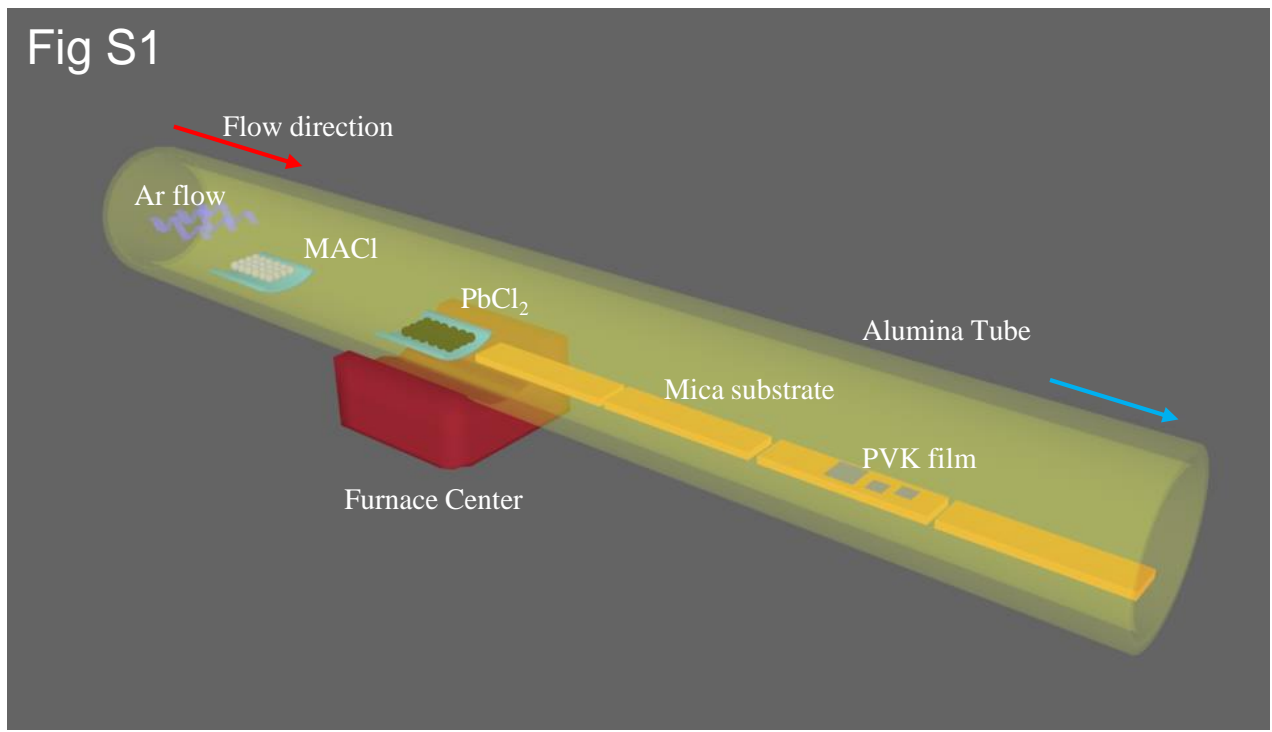

**Fig S1 Experimental setup of the CVD synthesis of MAPbCl$_3$.**

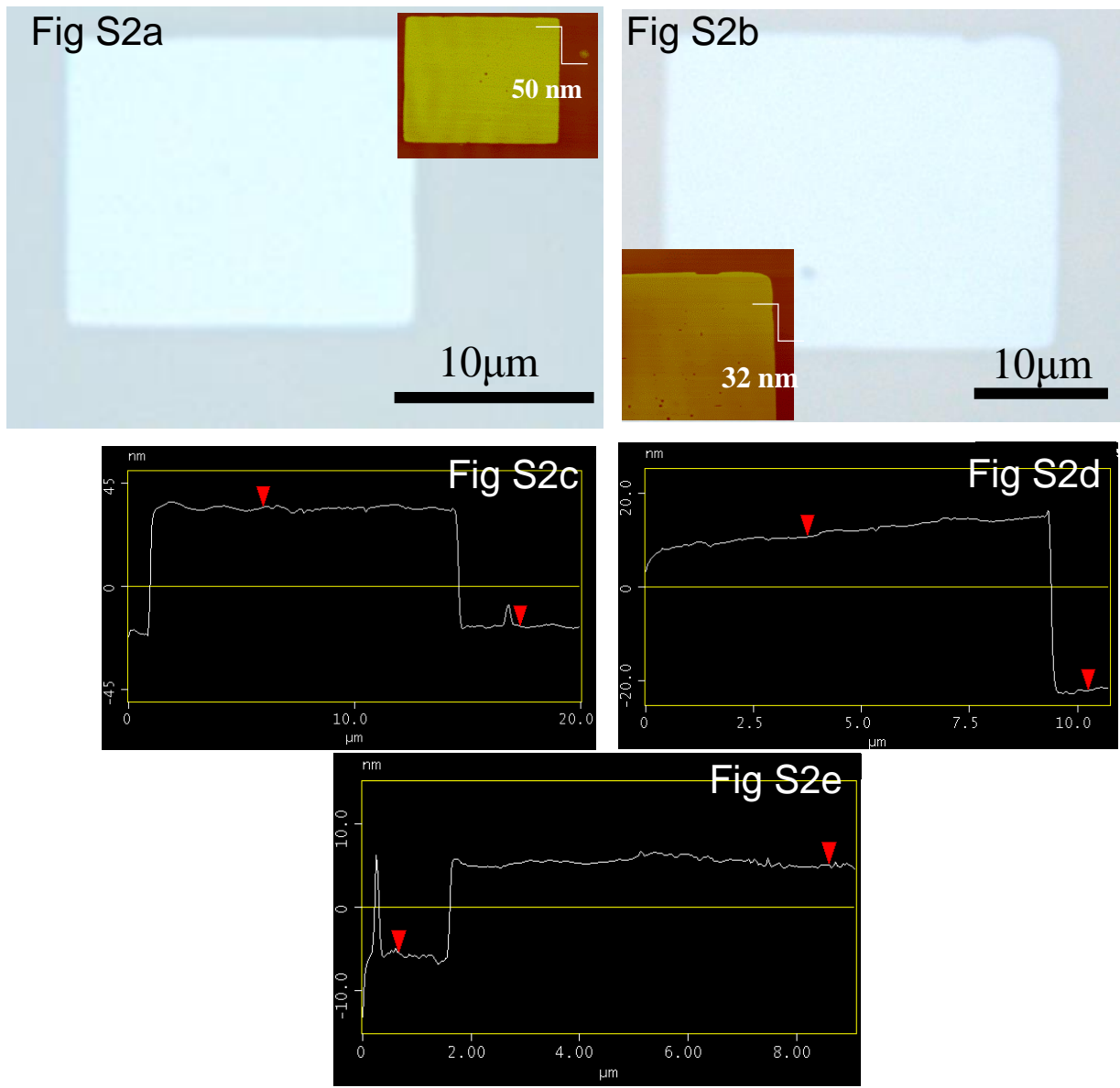

**Fig S2a-e Optical images and AFM results of Square PVK.** Square films that are more "whitish" in color show a larger thickness compared to the blue and opaque ones shown in the main text figures. Insets of Fig S2a and b show the AFM characterization-determined thickness for their respective films. Fig S2c-e show respectively the cross section analysis of the square films in Fig S2a, S2b and Fig 1(c) in the main text with x axis as the lateral distance and y axis the height in nm. All three section analysis reveals very flat thin film surface. The red triangles indicate the location where the thickness is measured.

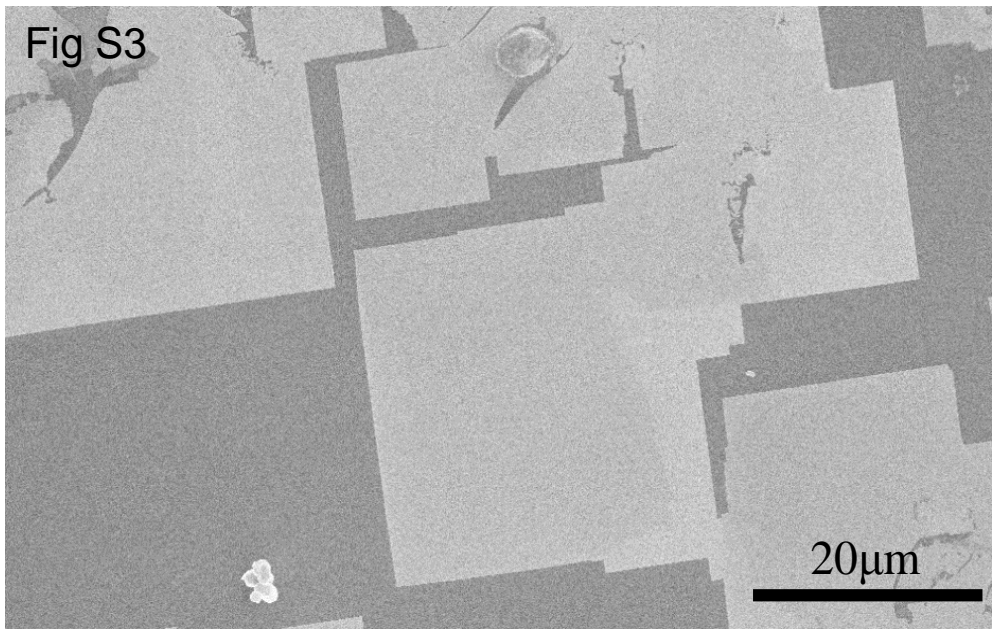

**Fig S3 Low magnification SEM image of Square PVK films.**

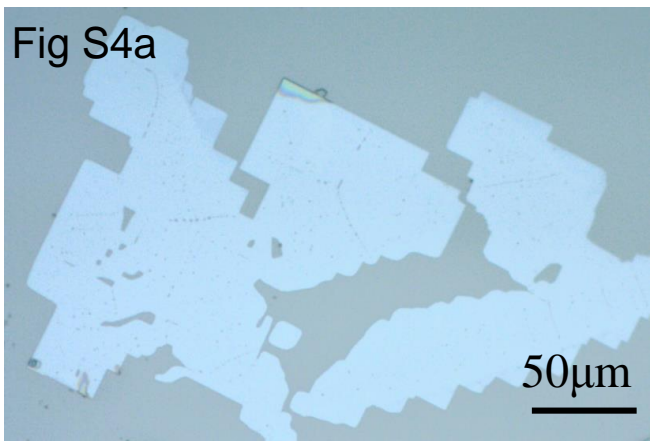 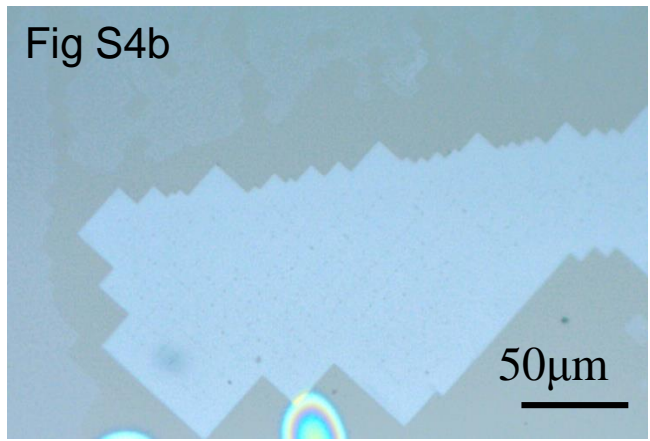

**Fig S4a, b Optical images of fractal PVK films.** PVK square films with the "stair-like" zig-zag morphology are commonly observed and they usually present themselves as very large scale, up to several hundreds of microns.

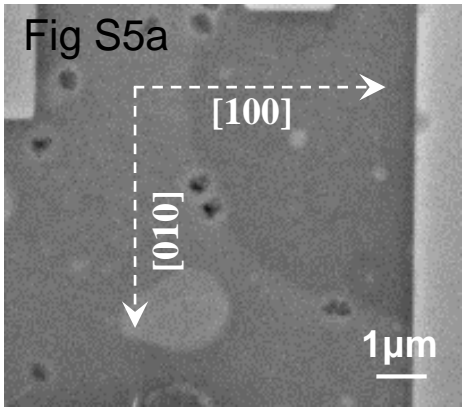 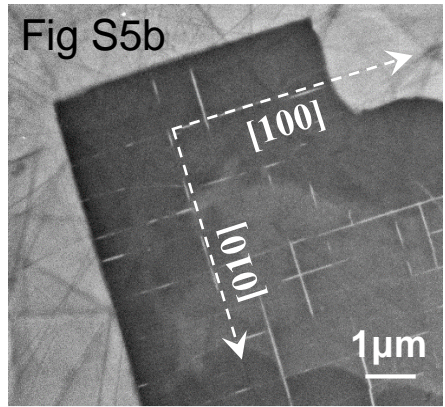

**Fig S5a, b TEM images of the as-transferred PVK film on copper grid.** Fig S5a shows the only presence of PVK film without mica substrate. The edges of the film correspond to the low-surface energy {100} facets. The PVK film is not stable under high energy electron beam which causes un-determined black dots on the film surface. In Fig S5b both Mica and PVK are present. Often additional crack lines (white line contrasts in Fig. S5b) along <100> directions are observed when PVK films stay on mica substrates. The crack lines are believed to be a consequence of the low-cohesive energy of the PVK material and the substrate-film strain induced during growth and TEM sample preparation processes. Based on recent progress on 2D materials VDW growth[5], a very small strain does exist between VDW film and VDW substrate. The magnitude of such strain is strongly correlated with the magnitude of VDW force, which is a few orders lower in magnitude than chemical epitaxial bonds. For materials with strong cohesive energy, the VDW strain should be quite small and negligible; while for materials with weak cohesive energy (e.g. $MAPbCl_3$), such VDW strain should be substantial. Since $MAPbCl_3$ has a very weak cohesive energy along <100> directions, the cracks we have observed are therefore a possible way for $MAPbCl_3$ to relax the weak VDW strain energy. It should be noted that once mica is removed the PVK film's crack lines disappear as shown in Fig. S5a, which is reasonable as crack lines representing high free energies are not thermodynamic stable during substrate-free scenario.

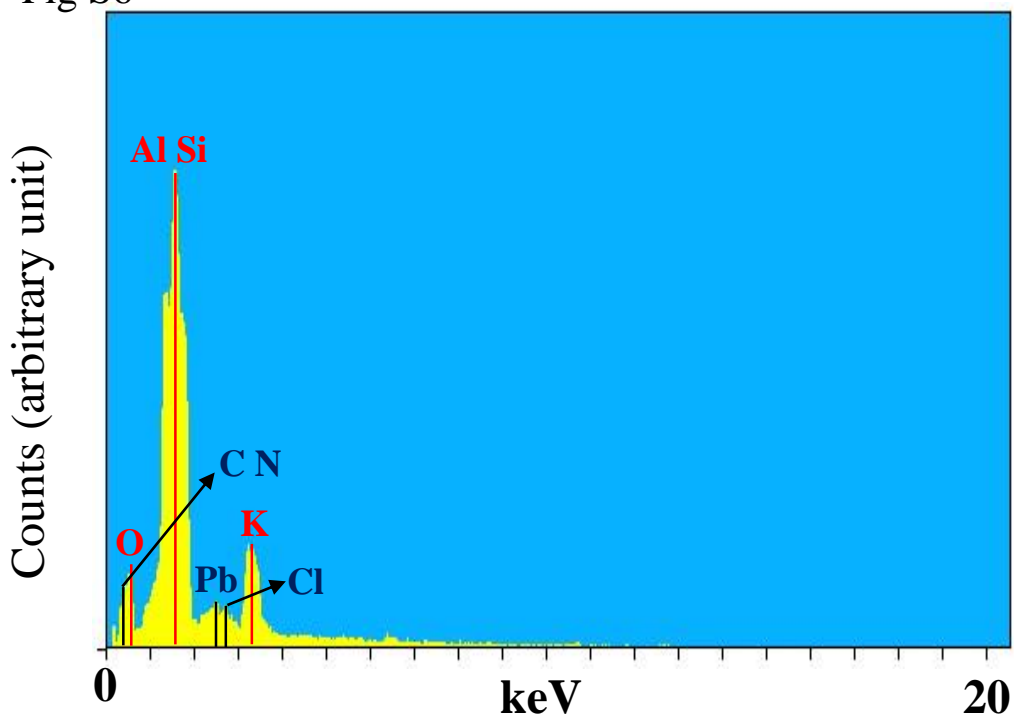

**Fig S6 EDS measurement of the as grown PVK film.** Several distinct peaks can be observed in the EDS measurement where the stronger ones marked by red correspond to mica substrate, a monoclinic structure composed of potassium and aluminosilicates. The relatively weak peak marked by black corresponds to the signatures of C, N, Cl and Pb elements, with the ones for Pb and Cl being more distinguishable. The chemical compositions of $MAPbCl_3$ include: C, N, H, Pb and Cl. H could not be detected by EDS. Both C and N always show up during most EDS characterizations due to the intrinsic hydrocarbon and nitrogen contamination in vacuum chamber. Therefore, the only accurate elements for EDS for $MAPbCl_3$ are Pb and Cl. We conclude the $MAPbCl_3$ structure and composition with several more evidences. First from the point of view of the crystal structure, only $MAPbCl_3$ possess a stable cubic structure while $PbCl_2$ is orthorhombic and shares a completely different lattice parameter (5.69Å vs. 4.5Å). The diffraction pattern in Fig 2(f) and (g) yields a lattice spacing of precisely 5.69Å, consistent with the $MAPbCl_3$ perovskite parameter. If any non-uniformity exists, it should be observed directly from the diffraction pattern. The second evidence for the confirmation of $MAPbCl_3$ is that we have already identified three typical results of the deposition, namely $PbCl_2$, $MAPbCl_3$ and MACl. Each of them has its own deposition location temperature and morphology. $PbCl_2$ shows itself as hexagonal pallets as can be seen later (in Fig S9f) and more importantly, they are only found on the second substrate, a region closer to the furnace center therefore being at a higher temperature. MACl on the other hand, is an organic compound that has very weak bonding and is always found on the substrate farthest from the furnace. They present themselves as very thick depositions without well-defined structure. The square sheets are found neither in the two locations mentioned above, but rather in between (the third substrate) where temperature is moderate and suitable only for the growth of $MAPbCl_3$. Since we never observed the coexistence of the two morphology of hexagonal pallets and square sheets, the possibility of non-uniformity in the composition can be ruled out.

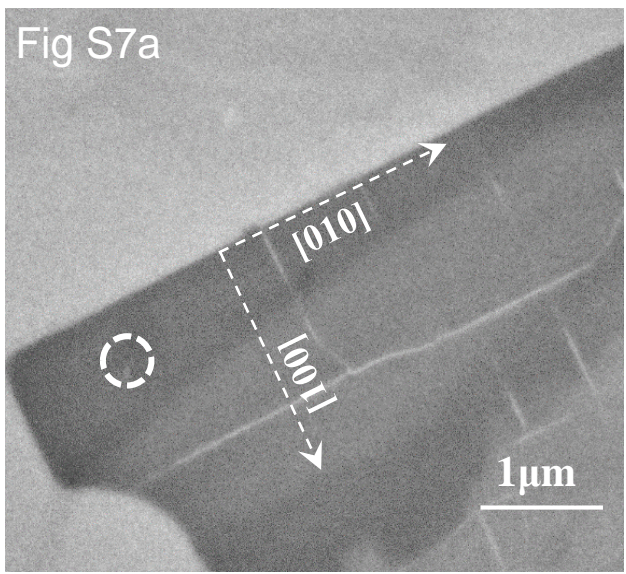 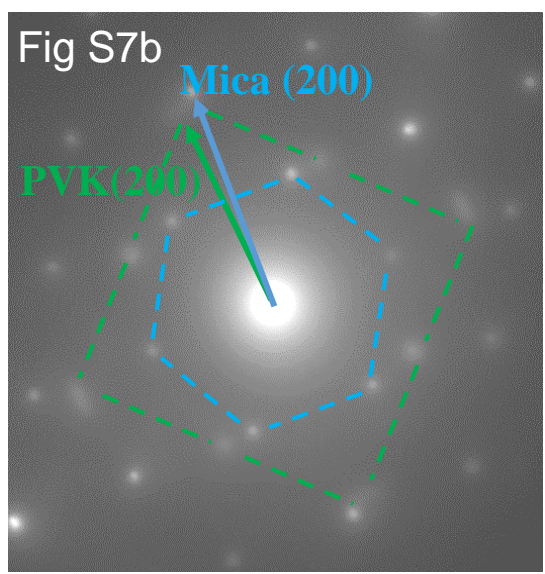

**Fig S7a, b TEM image and electron diffraction pattern of the PVK film on mica.** Fig S7a shows the TEM image of both mica (background) and PVK. Diffraction pattern (Fig S7b) was taken at the location marked by the white dashed circle in Fig. S7a. The epitaxial relation, as highlighted by the green (PVK) and blue (mica) features, is the same as Fig 2(g) in the main text.

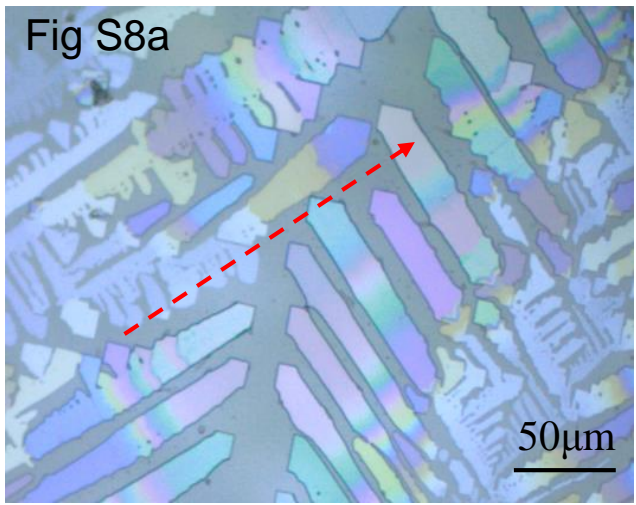 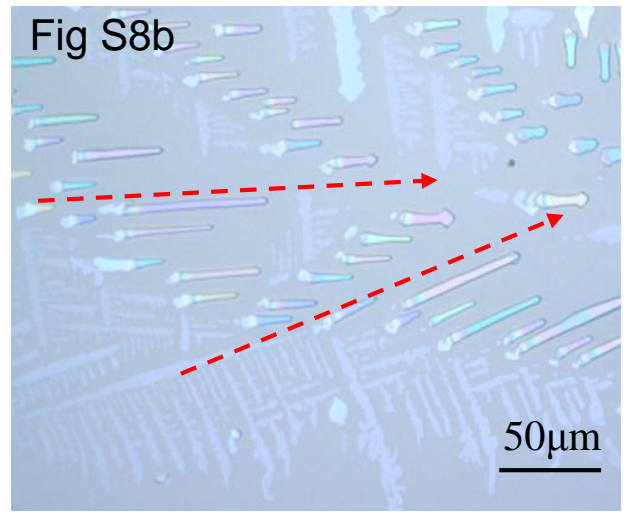

**Fig S8a, b Additional optical images of PVK dendritic films with different number of nuclei.** Dendritic morphology can tell nucleation condition by observing its orientation. Fig S8a shows the perpendicular dendrites that reflects only one nucleus; while in Fig S8b, two sets of dendrites with tilted orientations (marked by the red arrow) can be seen, indicating two initial nuclei.

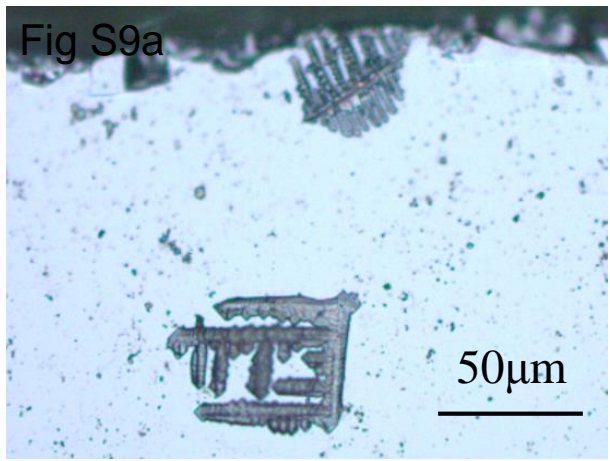 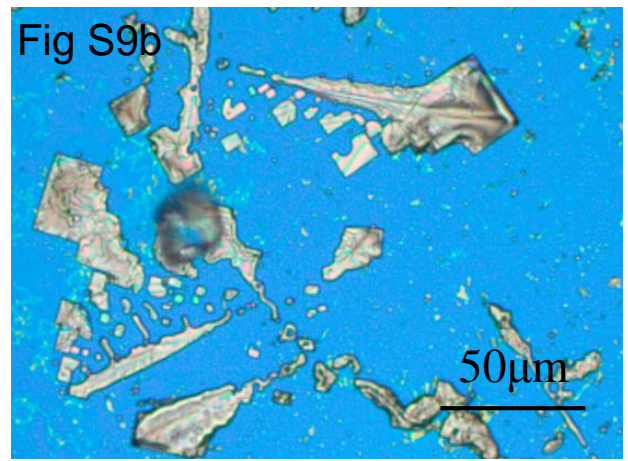
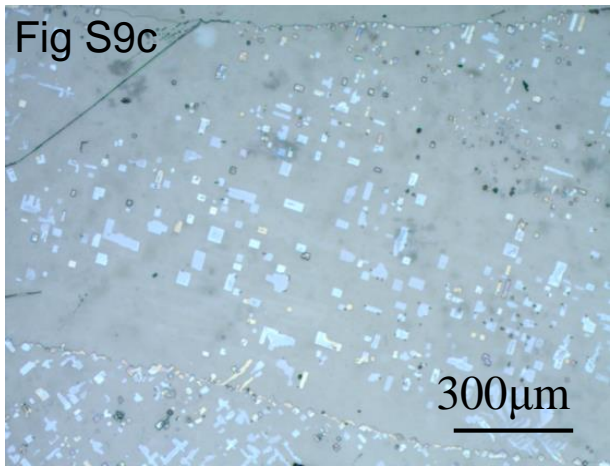 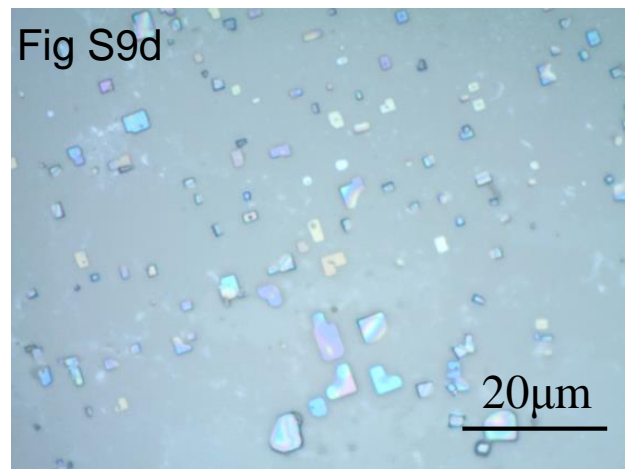
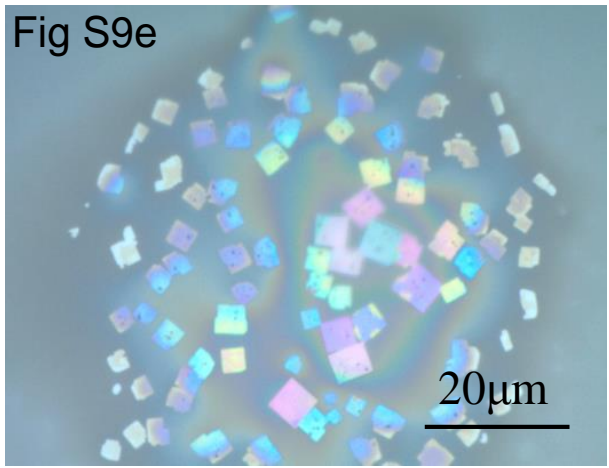 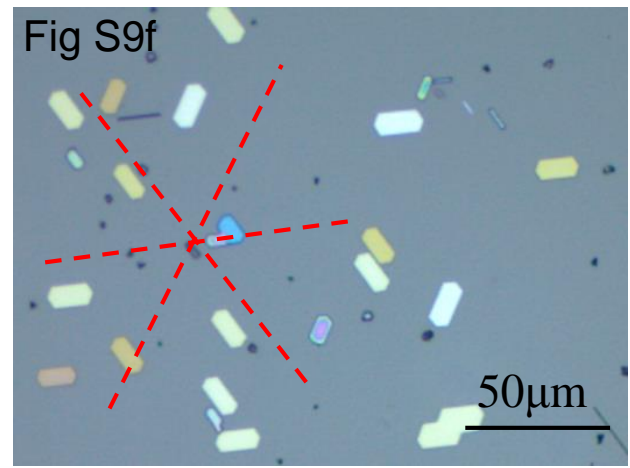

**Fig S9a-f Morphology of PVK films under different growth conditions:** PVK on HF-etched silicon wafer (Fig S9a); PVK on silicon wafer with graphene coated (Fig S9b); PVK on mica with graphene coated (Fig S9c); PVK on mica, the backside of Fig S9c (Fig S9d); PVK on mica with pre-deposited MACl (Fig S9e); $PbCl_2$ on mica (Fig S9f). See subsequent text for further discussion.

**Discussion on Fig S9a-f**

The HF-etched silicon wafer possesses a rough surface with dangling Si bonds and thus the growth falls into the conventional category. From Fig S9a we see large number of black dots present on the surface, indicating a very easy nucleation process and bulk morphology. Also present is two large bulk dendritic islands with perpendicular propagation direction which assembles those of VDW epitaxy on mica, only with very large thickness and in different orientations. The large thickness indicates very strong preference for vertical growth. Thinner films are obtained when the Si wafer has one layer of graphene transferred above (Fig S9b), since the adsorption energy ($E_{ad}$) is dramatically reduced. However, due to the fact that the polycrystalline graphene itself is transferred, instead of grown, onto the Si wafer, surface flatness cannot be assured and defects exist. The defects and grain boundaries would all serve as the preferred nucleation sites for PVK. The factors above make the 2D growth on graphene far worse than that on mica. Such a trend is preserved when graphene is transferred onto mica (Fig S9c), whereas on the other side of the same substrate (without graphene), the large scale and uniform 2D growth still exists (Fig S9d, notice the change in scale bar). Also interestingly, MACl, which has a much lower melting point is occasionally pre-deposited on the substrate farthest from the precursor (lowest temperature), most probably during the ramping process, as shown in Fig S9e the rainbow-like color contour beneath the square films. Subsequent deposition of PVK may occur on this pre-deposited layer, which theoretically should be classified as conventional growth. Indeed, in Fig. S94, multiple nuclei are indeed observed. Finally we used powdered $PbCl_2$ as a single precursor and the growth results are hexagonal films (Fig S9f), the same morphology as they appear in literature *(5)*. The absence of organic ions leaves only the Pb-Cl bond which is stronger and therefore making the nucleation process easier. As a result, we observed the multiple-nuclei here (as shown by the red dashed lines).

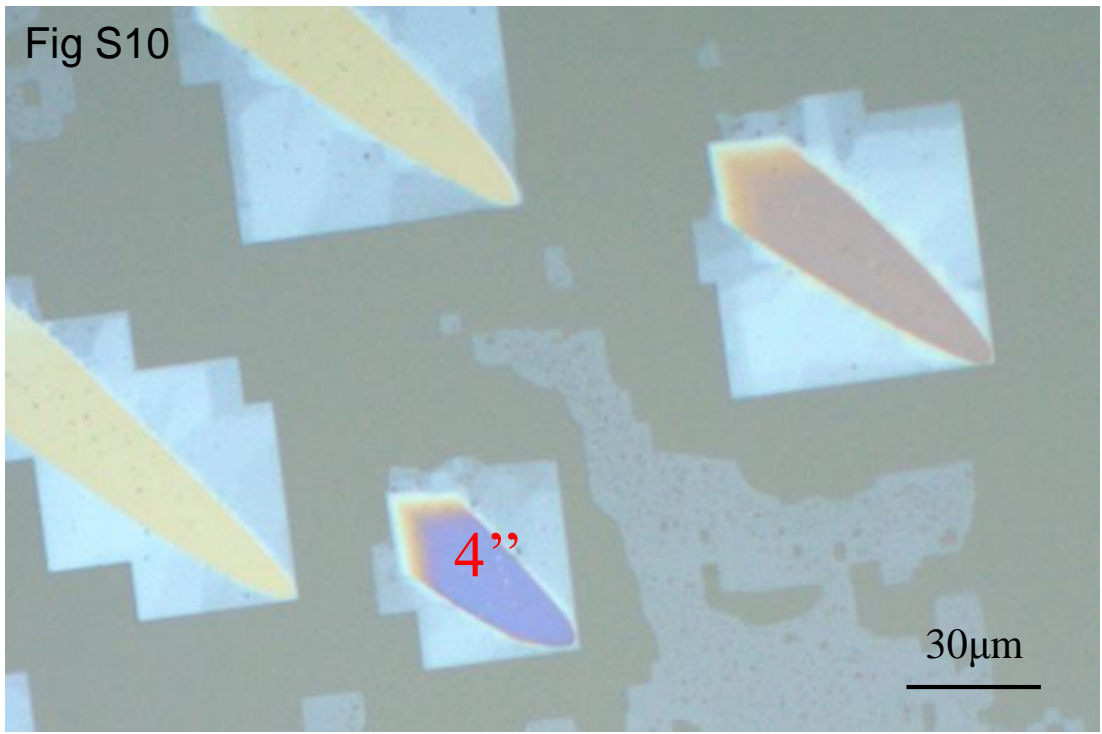

**Fig S10 Morphology evolution of PVK film.** It shows the morphology (location 4'' sample) of the sample at location 4' in Fig. 4(k) after one week at room temperature. From the location 4'' we see the zig-zag feature on the right-top corner of the square film at location 4' in Fig. 4(k) healed itself to be a complete square. The overall evolution of morphology follows the thermodynamic argument.

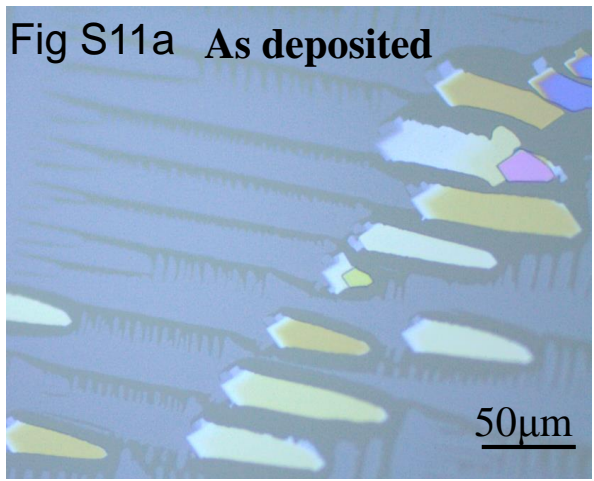 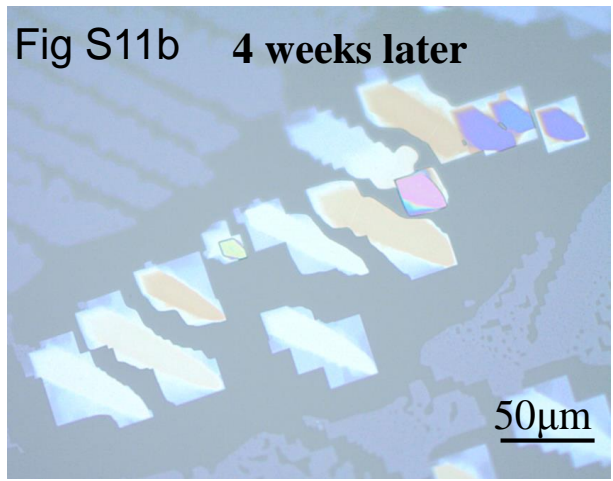

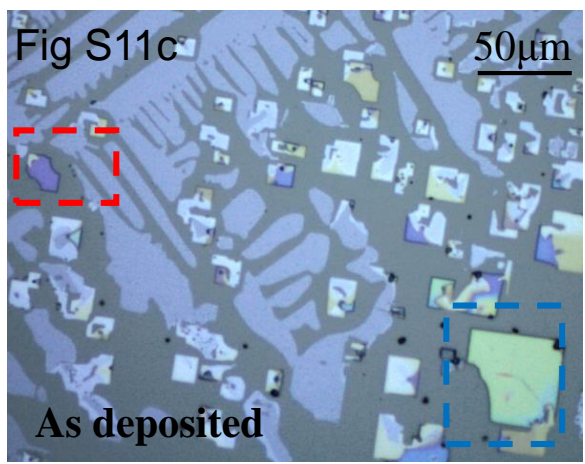 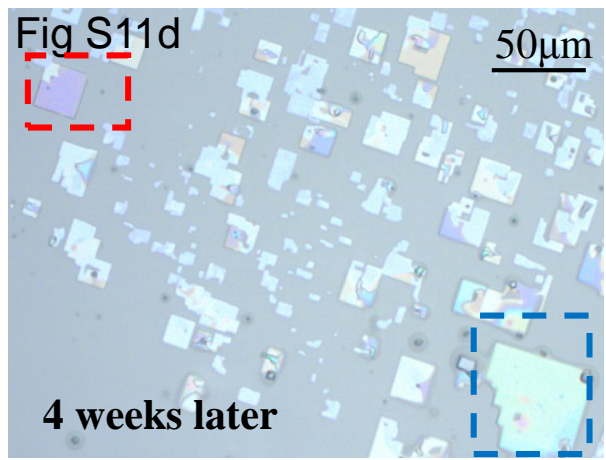

**Fig S11a-d Morphology evolutions of PVK films at room temperature.** Fig S11a-b show the similar morphology change as described in the main text where thicker film attracts more atoms and thinner ones disappear over time. Fig S11c-d show a similar but more dramatic change where the red dashed box-highlighted fractal film on the top-left corner completely healed itself to a square one with uniform thickness after aging. The blue box-highlighted film with smoother surface transformed to the fractal one with zig-zag thermodynamic stable facets after four weeks.

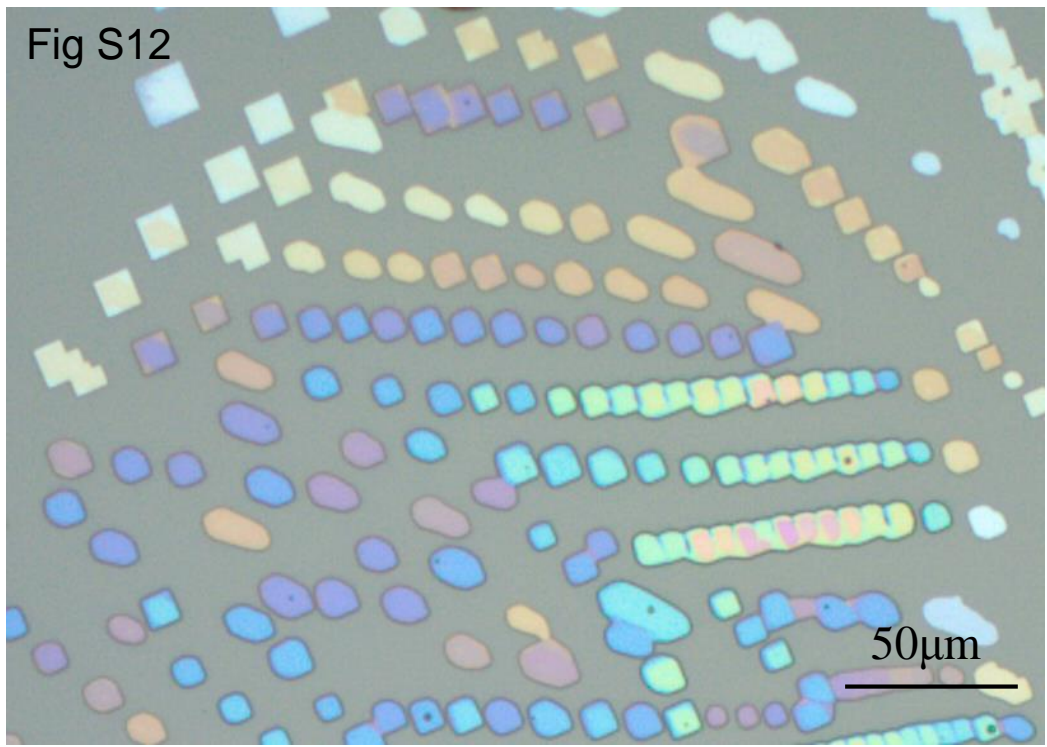

**Fig S12 Optical image of PVK films as a result of a relaxation process during deposition.** The relaxation process (i.e. a process driven by the surface free energy of PVK materials when adatoms' diffusivity are high) is more vigorous during deposition, evidenced by the presence of high density discrete films right after growth.

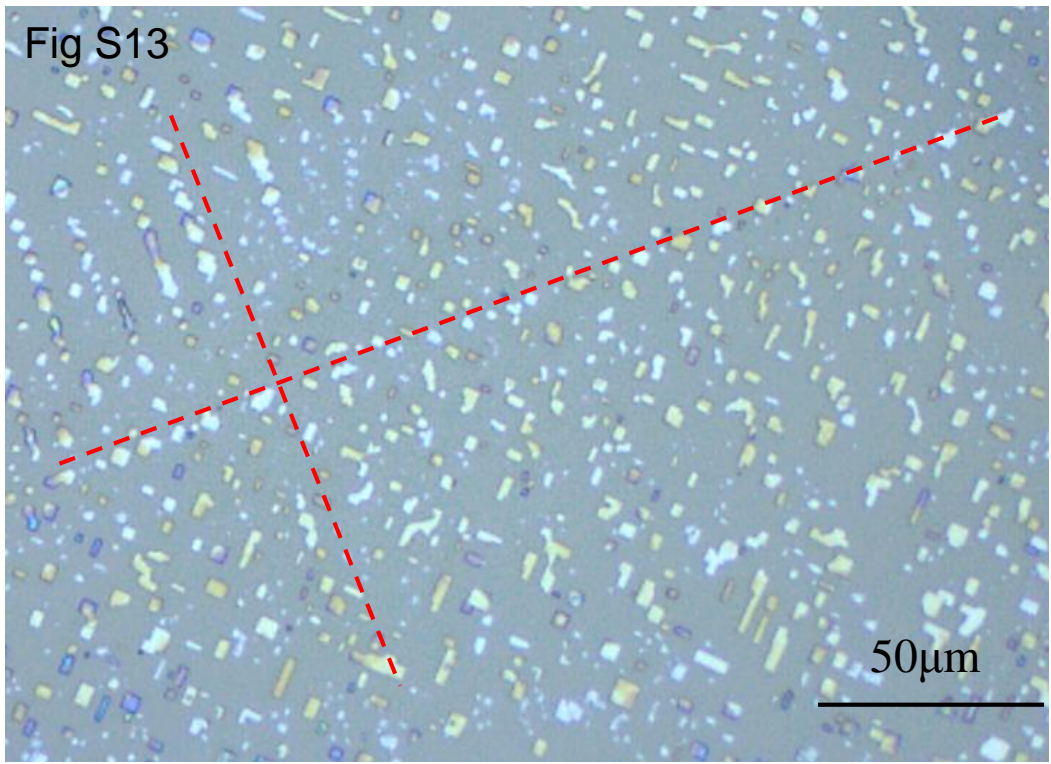

**Fig S13 Optical image of PVK films evidencing the relaxation process during deposition.** The periodicity of the distribution of square films (red dashed lines) indicates that all crystals stemmed from a single nucleus. The results were obtained right after growth without further aging.

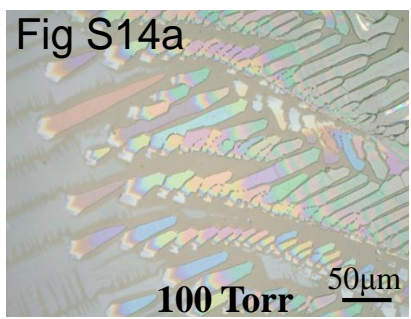
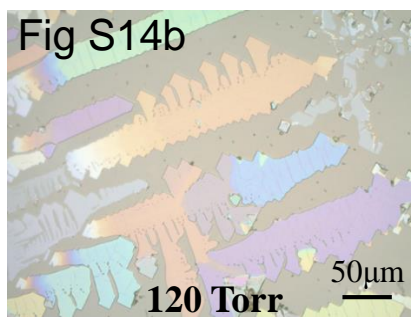
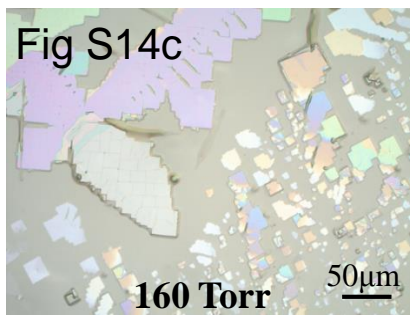
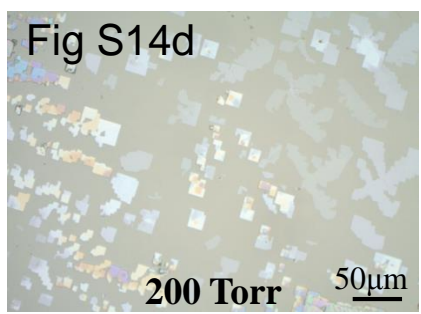
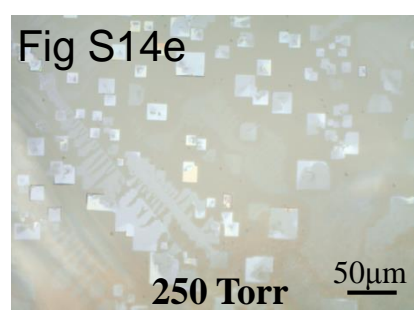

**Fig S14a-e Modulation of the relaxation process during deposition by chamber pressure.** At high deposition rate regime, increasing chamber pressure would reduce the overall deposition rate (more collisions are involved), which in turn favors the relaxation process. As it can be seen, with increasing chamber pressure (100-250 Torr), the PVK thin film morphology changes from dendrites dominated by individual main protrusions (Fig S14a) to dendrites with more branches (Fig S14b and c) and finally to more disconnected square films (Fig S14e and f).